\documentclass{article}

\usepackage[dvips]{graphics}
\begin{document}

\title{Dynamic nuclear Overhauser shifts in Larmor beats from a quantum well}%
\author{A.Malinowski and R.T.Harley\\
Department of Physics and Astronomy, \\Southampton University,
    Southampton SO17 1BJ,
    UK.}
\date{}%
% ----------------------------------------------------------------
\maketitle
\begin{abstract} The significance of nuclear spin
polarisation in time-resolved optical studies of III-V
semiconductors is addressed. Electron Larmor beats in pump-probe
reflectivity from a GaAs/AlGaAs quantum well show Overhauser shift
of 0.7 T due to accumulated nuclear polarisation
$\langle$I$\rangle$/I=0.065. This leads to precision values of
electron g-factor, elucidates nuclear spin pumping and diffusion
mechanisms in quantum wells and informs discussion of implications
for spin-electronics and transport.
\end{abstract}

% ----------------------------------------------------------------

The proposed application of electron spin transport and coherence
in semiconductors to spin electronics and perhaps quantum
computing \cite{kik99,hag98} will involve a high degree of
conduction electron spin polarisation and extended electron spin
lifetimes \cite{kik97}. Recently free Larmor precession continuing
for $\geq100$ ns \cite{kik98} and spin transport over microns
\cite{kik99,cam96} were observed in time-resolved studies of
n-type GaAs and quantum wells using circularly polarised optical
pulses. On the other hand, since all naturally occurring isotopes
in III-V semiconductors have non-zero nuclear spin \cite{shi96},
hyperfine interaction A$\mathbf{I.S}$, can be expected to result
in an equivalent level of polarisation of the nuclear spins
\cite{mei84} with, conversely, profound effects on the behaviour
of the electrons. These  include built-in effective magnetic
fields (Overhauser effects) and nonlinear or bistable response to
applied external fields \cite{mei84}. Such effects have long been
known in cw optical phenomena in bulk semiconductors \cite{mei84}
and in quantum wells \cite{sne91,eki72} and in quantum
magneto-transport \cite{dob88}, but consequences for time-resolved
optical measurements and spin applications have been neglected. In
this letter we report measurements of nuclear Overhauser shift
\cite{mei84} in ps time-resolved Larmor beats \cite{heb94,ama97}
in a GaAs/AlGaAs quantum well, revealing the dynamics and long
timescales of nuclear spin polarisation by optical pumping and the
change of electron spin precession frequency induced by a
polarised nuclear spin system. In this experiment the effective
internal field approaches 1 T although the nuclear polarisation is
only 6.5\%, emphasising the significance of nuclear phenomena.
Proper account of the Overhauser shift gives excellent agreement
between measured conduction electron Larmor frequencies and
anisotropic k.p theory \cite{ivc92}, while the dynamics provides
new insights to nuclear spin diffusion in quantum well systems. We
suggest that such effects may commonly influence and could be
fruitfully exploited in time-resolved polarised measurements.

In our experiment (see inset to figure 1) a circularly polarised
optical pump pulse from a mode-locked laser is absorbed in the
quantum well (QW) at time $t_{0}$ and generates a transient
population of conduction electrons with spin-polarisation
$\mathbf{S}$($t_{0}$), parallel (antiparallel) to the incident
beam for $\sigma-$ ($\sigma+$) polarisation \cite{mei84}.
Time-evolution of $\mathbf{S}$ is monitored through rotation, on
reflection, of the plane of polarisation of a weak, delayed probe
pulse, almost colinear with the pump, the rotation being
proportional to the population difference of the electron
spin-states \cite{bri98}. Magnetic field, $\mathbf{B}_{ext}$,
applied at angle $\theta$ to the beam, causes the spins to precess
while $\mathbf{S}(t)$ decays comparatively slowly  by
spin-relaxation and recombination so that there is a
non-precessing component of electron spin, $\mathbf{S}_{av}$,
parallel to the field. Integrated over many laser pulses, nuclear
spin polarisation, $\langle$I$\rangle$, parallel to
$\mathbf{S}_{av}$, will build-up through mutual spin flip-flops
with lattice nuclei driven by the hyperfine interaction.  The
polarisation $\langle$I$\rangle$ reacts back on the electron spins
as an effective (Overhauser) magnetic field \cite{mei84},

\begin{equation}
\mathbf{B}_{N}=\frac{A\langle\mathbf{I}\rangle}{g_{e}\beta}
\approx\frac{\langle \mathbf{I} \rangle}{I}\mathrm{tesla}
\label{eqn1}
\end{equation}

where $g_{e}$ is the electron g-factor and $\beta$ is the Bohr
magneton. The numerical value is calculated for a 9.6nm GaAs
quantum well with $\theta$=45$^{o}$ (see below), using an average
of $A$ over the isotopes  $^{75}$As,  $^{69}$Ga and  $^{71}$Ga
each with I=3/2 \cite{mei84}. This is a significant field even
when the nuclear polarisation is low. When ${g}_{e}$ is negative,
as for bulk GaAs and GaAs/AlGaAs quantum wells wider than 5.5 nm
[8], $\mathbf{B}_{N}$ opposes $\mathbf{S}_{av}$ (see fig. 1) and
the Larmor frequency is $\Omega$ =
$g_{e}\beta$($\mathbf{B}_{ext}+\mathbf{B}_{N}$)/$\hbar$. Reversal
of the polarisation of the pump reverses $\mathbf{S}_{av}$ and
therefore also $\langle$I$\rangle$ and $\mathbf{B}_{N}$, so
increasing the Larmor frequency by 2$\mid
g_{L}\beta\mathbf{B}_{N}\mid$/$\hbar$.

For small nuclear polarisation  and for
$\mathbf{B}_{ext}\gg\mathbf{B}_{L}$ the internuclear dipolar field
(~10$^{-4}$T), the steady state value is \cite{mei84}

\begin{equation}
\frac{\langle\mathbf{I}\rangle}{I}=\frac{5}{3}\left(\frac{\tau_{s}}
{\tau_{s}+\tau_{r}}\right)\left(\frac{T_{1}^{*}}{T_{1}^{*}+T_{1e}}\right)
\left(\frac{\mathbf{S}(t_{0})}{\mathbf{S}}\right)\cos\theta
\label{eqn2}
\end{equation}

where  $t_{s}$ and $t_{r}$ are respectively electron
spin-relaxation and recombination times;  $T_{1}^{*}$ and $T_{1e}$
are respectively times characterising nuclear spin relaxation and
angular momentum transfer from electronic to nuclear spins. As we
shall see, the latter may involve spin diffusion as well as
hyperfine coupling. We neglect the equilibrium electron spin due
to applied field, which is of order 1\% in the highest fields used
in this experiment.  Equation 2 shows that accumulation of nuclear
polarisation can be  avoided if $\theta$=90$^{o}$. The electron
spins then precess on a disc, not a cone. $\langle$I$\rangle$ is
also zero if $\mathbf{B}_{ext}\ll\mathbf{B}_{L}$ \cite{mei84},
often experimentally impractical. The main conditions, readily
met, for build-up of $\langle$I$\rangle$ are $t_{s}\geq t_{r}$ so
thatphotoelectrons remain spin-polarised during their lifetime,
and  $T_{1e}\leq T_{1}^{*}$ so that angular momentum transferred
from electron to nuclear spin systems is not rapidly dissipated to
the lattice. In III-V semiconductors, values of  $t_{s}$ vary from
$\sim$5 ps to $\sim$100 ns
\cite{kik98,mei84,bri98,mai93,har94,wag93,tac97} while typically
$t_{r}\geq$ 100 ps. For Bloch states,  $T_{1e}$ has been estimated
as 10$^{4}$ s in GaAs for excitation density ~10$^{15}$ cm$^{-3}$
\cite{mei84} but it is reduced by a factor ~10$^{5}$ for nuclei in
contact with electrons localised, for example, by interface
roughness or donors \cite{mei84}. The nuclear spin-lattice
relaxation time is of order minutes or hours at low temperatures,
driven by phonon-modulation of electric field gradient in
intrinsic material \cite{abr61} or mediated by hyperfine coupling
to electron spins in n-type systems \cite{ber90}.

Our sample consisted of three undoped single GaAs quantum wells of
widths 5.1 nm, 9.6 nm and 19 nm with 34 nm Al$_{.33}$Ga$_{.67}$As
barriers and grown on an (001)-oriented n+ (10$^{18}$ cm$^{-3}$
Si-doped) substrate. A bias applied between the substrate and an
ITO electrode on the top of the structure controlled the carrier
type and concentration in the wells. Most of the measurements
described here were performed at 10 K on the 9.6 nm well with
0.5x10$^{11}$ cm$^{-2}$ heavy holes injected to ensure that the
effects were dominated by the photoexcited electrons \cite{ama97}.
In fact, biasing the sample flat-band was found to make no
significant difference to the results. This structure was also
chosen to allow tests of inter-well spin diffusion. Pump and probe
pulses were derived from a mode-locked Ti-Sapphire laser giving 2
ps pulses at 80 MHz repetition frequency and tuned to the n=1
heavy-hole absorption edge \cite{bri98}. We estimate pump
excitation density ~1.0x10$^{10}$ cm$^{-2}$ per pulse in the
quantum well. The electron lifetime and spin relaxation time were
determined to be $t_{r}\approx$240 ps \cite{sne91} and
$t_{s}\approx$200 ps respectively. In the actual experimental
geometry, dictated by the configuration of the magnet, the beams
intersected at 3$^{o}$ on the sample, at 45$^{o}$ to the quantum
well normal giving propagation close to the normal inside the
sample, due to the high refractive index. The pump-induced probe
polarisation rotation (typically $\leq 0.1^{o}$) was measured by a
sum-frequency lock-in technique \cite{bri98}. Magnetic field,
$\mathbf{B}_{ext}$, was applied perpendicular to the incident
beams, at  45$^{o}$ to the normal to the quantum wells.

Data for linear pump polarisation, fig.1(a), i.e. nominally equal
electron spin populations, shows a background resulting from
nonidealities of the experimental arrangement. Circular pump
polarisation, fig.1(b), gives Larmor beating, which decays due to
electron spin-relaxation and recombination. The spin vector,
initially parallel to the pump beam, precesses on a cone of
half-angle $\theta\sim$ 45$^{o}$ (see inset) and, after an odd
number of half periods is perpendicular to the pump, corresponding
to an equal coherent superposition of spin-states and giving zero
probe polarisation rotation. After an even number of half periods
the spin state is pure $\sigma +$ or $\sigma -$, so rotation
should be a maximum absolute value \cite{ama97}. The observed
beats are consistent with this expected behaviour, superimposed on
the background of fig. 1(a). Solid curves in fig. 1(b) are
numerical fits used to obtain the Larmor frequency.

A frequency shift between the Larmor beats for the two pump
polarisations, is clearly apparent in fig. 1(b). The only change
between the traces is that of pump polarisation, unambiguously
demonstrating that the effect is due to an internal field
associated with the electron spin orientation and ruling out any
interpretation in terms of stray applied fields. The shift
corresponds to an Overhauser field $\mathbf{B}_{N}\sim$0.2 T
opposite to $\mathbf{B}_{ext}$ for $\sigma-$ pump and hence to
negative g$_{e}$.

The data of fig. 1 were taken after ~100 s of exposure to pump
light. Repeating the experiment after longer exposures to pump
light showed a slow time evolution in the Larmor frequency,
corresponding to a slow increase of $\mathbf{B}_{N}$ to a
saturation value $\mathbf{B}_{N}$= 0.7 T after about 3000 s,
corresponding to $\langle$I$\rangle$/I= 0.065 (eqn.1). Following
saturation, with the beams blocked the value of $\mathbf{B}_{N}$
fell by 30\% over a period 600 s, giving  $T_{1}^{*}\gg$ 1.8
x10$^{3}$s. Inserting this and the known values of  $t_{s}$ and
$t_{r}$ into eqn. 2 and assuming that S$(t_{0})$/S$\sim$0.8, we
find $T_{1e}$~10$^{4}$ s. This is comparable to values calculated
for hyperfine coupling to Bloch electrons \cite{mei84}, but, as
discussed below, is almost certainly a value associated with
polarisation by localised electron states and limited by nuclear
spin diffusion.

Figure 2(a) shows Larmor frequencies for $\sigma-$ pump
polarisation as a function of $\mathbf{B}_{ext}$ after 100 s of
pumping (open circles) and with $\mathbf{B}_{N}$ saturated after
4000 s of pumping for each point (filled circles). The fitted
lines give $\mid\mathbf{B}_{N}\mid$=0.2 T after 100s and
$\mid\mathbf{B}_{N}\mid$=0.71$\pm$0.05 T after 4000s and
$g_{e}$($\theta$= 45$^{o}$)= -0.213$\pm$0.004. The triangles are
data for $\theta$= 90$^{o}$ and the dashed fit yields
$\mid\mathbf{B}_{N}\mid=0.04\pm$0.03 T and $g_{e}$($\theta$=
90$^{o}$)=-0.178$\pm$0.001. From these values we calculate [13]
${g}_{e}$($\theta$=0)=-0.243$\pm$0.004 for $\mathbf{B}_{ext}$
along the growth direction. Calculations based on k.p theory by
Ivchenko and Kiselev \cite{ivc92} give $g_{e}$($\theta$=
90$^{o}$)=-0.166 and $g_{e}$($\theta$=0)=-0.248 for 9.6 nm wide
quantum wells, in close quantitative agreement with the
measurements.

Figure 2(b) shows a measurement tracking the approach to
saturation for $\sigma+$ pumping, at $\mathbf{B}_{ext}$=4T, by
setting the pump-probe delay to 129 ps, a steeply changing portion
of the signal in fig.1(b) (The sign of the signal has been
reversed with respect to fig.1 for clarity). The initial rise
includes the switch-on of the signal for $\mathbf{B}_{N}$=0 and
the establishment of the ''instantaneous'' part of
$\mathbf{B}_{N}$ within 10 s. As $\mathbf{B}_{N}$ increases, the
Larmor frequency increases, and the signal observed at a fixed
pump-probe delay changes according to the gradient of the Larmor
oscillations at that delay. The observed change in signal yields a
time-constant for the buildup of $\mathbf{B}_{N}$ of
$T_{B}\sim$900 s. Combining this data with the determination of
$\langle$I$\rangle$/I after $\sim$100 s and after saturation as
described above, we can plot the form of pumping dynamics shown in
fig. 2(c). The two-timescale behaviour can be assigned to effects
of nuclear spin diffusion. The hyperfine interaction rapidly
polarises nuclei near to electron localisation centres
\cite{mei84} giving the ''instantaneous'' component of
$\langle$I$\rangle$/I$\sim$0.017. The local nuclear polarisation
then spreads into the intervening regions by diffusion, to build
up additional polarisation at a slower rate. This rate, neglecting
relaxation of the nuclear spins, is given by $T_{B}^{-1}\sim
Dn_{loc}$ where $n_{loc}$ is the concentration of localised
electronic states and D is the nuclear spin diffusion coefficient.
Taking D=10$^{-13}$ cm$^{2}$s$^{-1}$, measured for As in GaAs
[22], and our value of $T_{B}$=900 s, we obtain
$n_{loc}\sim$10$^{10}$ cm$^{-2}$.

Establishment of equilibrium involves in-well spin-diffusion over
$\sim$100 nm and if the diffusion coefficient in the 34 nm
barriers were similar there would be significant inter-well spin
transfer. We investigated diffusion across the barriers by, first,
optically pumping the 19 nm quantum well in the sample for ~6000 s
to produce saturation of $\mathbf{B}_{N}$. This required tuning
the laser to the band edge for the 19 nm well, a photon energy not
absorbed in, so not directly pumping, the 9.6 nm well. The laser
beams were then blocked, retuned to the band edge of the 9.6 nm
well and unblocked to make a rapid measurement of $\mathbf{B}_{N}$
for that well; this gave only the 'instantaneous' value. Therefore
little, if any, nuclear polarisation had diffused through the 34
nm AlGaAs barrier during prolonged pumping of the 19 nm well. An
upper limit for spin-diffusion coefficient in the barrier  is
therefore ~1x10$^{-14}$ cm$^{2}$s$^{-1}$ (assuming the dark
relaxation time, $T_{1}^{*}$ is not significantly lower in the
barriers). This finding complements those from optically pumped
NMR \cite{lam68}, which show $^{71}$Ga signal build-up from nuclei
in the barrier over $\sim$200 s, due to (limited) spin-diffusion
into the barriers. Spin diffusion occurs mainly via mutual spin
flips of like isotopes which conserve the Zeeman energy of the
spin system. Therefore inter-well diffusion in GaAs/AlGaAs should,
indeed, be much slower than diffusion within the wells because
$^{27}$Al (I=5/2) gives magnetic spin disorder in the Ga
sublattice and also introduces disorder in the As sublattice due
to random quadrupole splittings.

In conclusion our results show the importance of nuclear spins in
time-resolved spectroscopy of III-V semiconductors. Among possible
applications we find that it is essential to take account of such
polarisation in measurements of electron g-factors and that new
insights are obtained into nuclear  spin diffusion in quantum
structures. The time-resolved  reflection technique allows
investigations in situations  inaccessible to traditional cw
measurements \cite{mei84,sne91} where  luminescence may be
short-lived compared to spin memory. In our GaAs/AlGaAs sample,
only 6.5\% nuclear spin polarisation is generated in the quantum
well, causing an effective internal field approaching 1 T, limited
by the comparatively short spin relaxation time of the electrons.
Much greater nuclear polarisations, beyond this essentially linear
(low-polarisation regime) should occur in weakly n-type GaAs under
strong pumping conditions, because of the long spin relaxation
times \cite{kik98} and donors which will enhance electron-nuclear
spin transfer. We note that it is just this type of material which
is under consideration for future applications in spin-devices.

We wish to thank Professors J.J.Baumberg and G.Bowden for
illuminating discussions.

% ----------------------------------------------------------------
\bibliographystyle{unsrt}

\setlength{\unitlength}{1cm}
\begin{center}
\begin{figure}
\begin{picture}(10,15)
\includegraphics{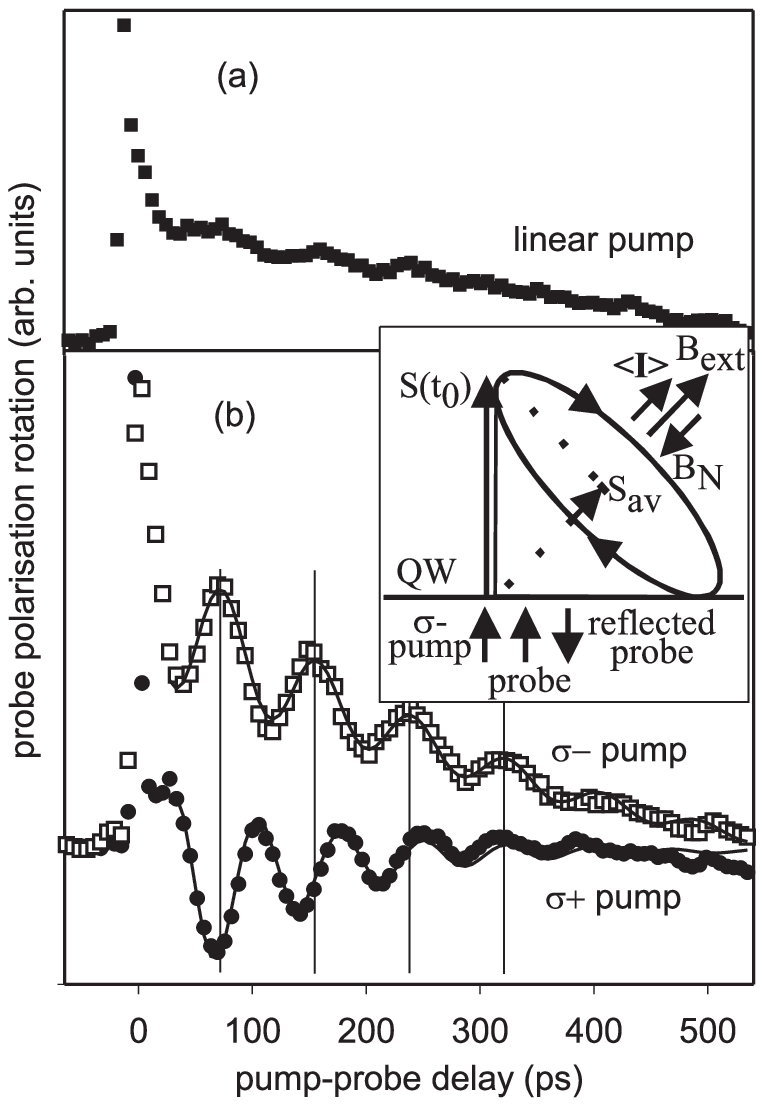}
\end{picture}
\caption{Representative results for pump-induced probe
polarisation rotation at 4 T in 9.6 nm single GaAs/AlGaAs quantum
well containing ~0.5x10$^{11}$ cm$^{-2}$ heavy holes at 10 K. (a)
Signal for linearly polarised pump shows background associated
with nonideality of experimental setup; (b) signals for circularly
polarised pump, showing Larmor beats and frequency shift due to
Overhauser field following about 100 s of exposure to pump light.
Inset shows principle of optical pumping of nuclear spin
polarisation $\langle$I$\rangle$ and associated Overhauser field
$\mathbf{B}_{N}$ in a quantum well (QW). }
\end{figure}

\begin{figure}
\begin{picture}(10,15)
\includegraphics{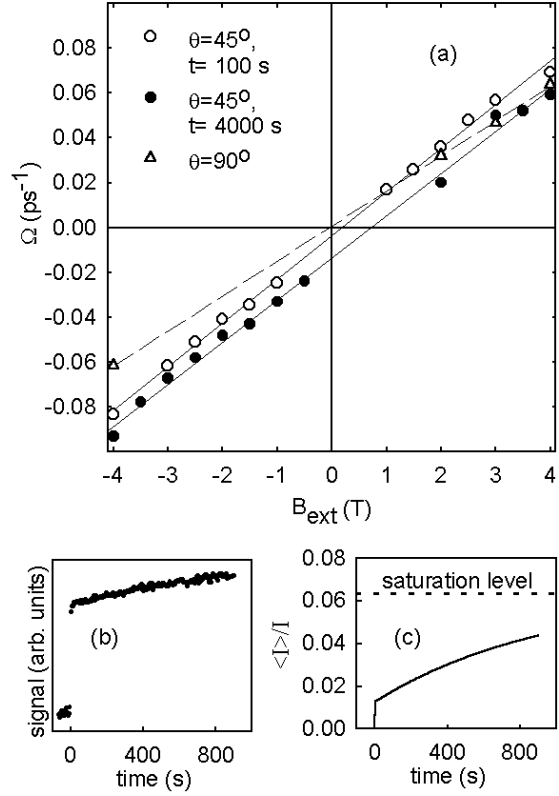}
\end{picture}
\caption{(a) Larmor frequencies observed for $\sigma-$ polarised
pump. Open circles and filled circles are for $\theta$=45$^{o}$
with pumping times 100 s and 4000 s respectively. Triangles are
for $\theta$=90$^{o}$ and pumping time 100 s. The solid and dashed
lines are fits giving $\mathbf{B}_{N}$ and ${g}_{e}$($\theta$)
(see text). (b) Time-dependence of Larmor beat signal in fig. 1 at
129 ps delay, midway between turning points, illustrating the
approach of the Overhauser field to a saturation value with a
time-constant T$_{B}\sim$900 s. (c) Inferred time-evolution of
nuclear spin polarisation.}
\end{figure}
\end{center}

\end{document}